\documentstyle[prl,aps,epsfig,multicol,amssymb]{revtex}
\begin{document}
\draft
\begin{multicols}{2}
{\bf \noindent ``Fluid-solid phase-separation in hard-sphere mixtures
is unrelated to bond-percolation.''
} \vspace{10pt}

In a recent letter, A.~Buhot\cite{Buho99} proposes that entropy driven
phase-separation in hard-core binary mixtures is directly related to a
bond-percolation transition.  In particular, Buhot suggests that a
phase-instability occurs when the coordination number $n_b$, defined
as:
\begin{equation}\label{eq1}
n_b = \rho_l \int_{\sigma_l \leq r \leq \sigma_l(1+R)} g_{ll}(r) d\bf{r},
\end{equation}
is equal to $z p_c$, where z is the coordination number of a
particular crystal lattice, and $p_c$ is its bond-percolation
threshold. Here $\rho_l$ is the number density of the larger
particles, $g_{ll}( r)$ is the radial distribution function of the
larger particles, and $R=\sigma_s/\sigma_l<1$ is the ratio of the
diameters $\sigma_i$.  However,  for binary
hard-sphere mixtures, calculations based on an accurate approximation
to $g_{ll}(r)$ demonstrate that $n_b$ varies widely along the
phase-boundaries calculated directly by simulations, implying that
bond-percolation is {\em unrelated} to the phase-separation in these
systems.

\begin{figure}
\begin{center}
\epsfig{figure=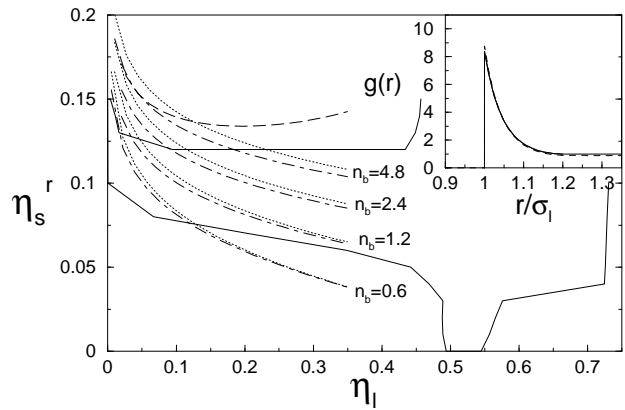,width=8cm}
\begin{minipage}{8cm}
\caption{\label{Fig1} Solid lines: the fluid-solid and fluid-fluid
phase lines from the simulations of Dijkstra {\em
et.~al.}\protect\cite{Dijk98} for $R=1/30$. Dotted lines and
dash-dotted lines: the coordination number calculated as in the text
using the AO potential and the potential from the simulations
respectively. 
Long-dashed line: the $n_b=2.4$ line  proposed by
Buhot\protect\cite{Buho99}.  {\bf Inset}: $g_{ll}(r)$ for $R=0.2,
\eta_l=0.25,\eta_s^r=0.25$ with AO potential. Solid line: from
$g_{ll}(r) = \exp(-\beta V_{dep}(r))$; dashed line: from PY integral equation.}
\end{minipage}
\end{center}
\end{figure}

For highly asymmetric binary hard-sphere systems, Dijkstra {\em
et.~al.}\cite{Dijk98} conclusively demonstrated that an effective
one-component description based on a depletion potential picture
quantitatively describes the fluid-solid transition. This in turn
implies that the one-component description should give a fair
representation of the radial distribution function $g_{ll}(r)$. Recent
simulations\cite{Dijk99} of the Asakura-Oosawa (AO) depletion
potential\cite{Asak54} show that the Percus-Yevick (PY) approximation
quantitatively describes the pair-correlations along the fluid-solid
transition line.  In the inset of Fig.~\ref{Fig1}, the simple form:
$g_{ll}(r) = \exp \left( - \beta V_{dep}(r) \right)$, where
$V_{dep}(r)$ is the effective depletion potential, is compared to more
the accurate PY integral equation results.  Typically for packing
fraction $\eta_l = \pi \rho_l \sigma_l^3/6 \leq 0.25$ along the phase
boundaries this form gives near quantitative agreement for $r \leq
\sigma_l(1+R)$, which is not surprising since for small $\eta_l$ the potential
is typically at least $2.5 k_B T$ along the fluid-solid transition
line while the hard-core induced correlations are small so that the
exponential form dominates.  In fact, Buhot's treatment of binary
hard-spheres reduces exactly to this simple form but with an AO
depletion potential which is valid only when $\eta_l \rightarrow
0$. If one replaces $\eta_s$ with the $\eta_s^r$ of small spheres in a
reservoir kept at constant chemical potential, the correct form of the
AO potential is recovered\cite{Dijk98}. In Fig.~\ref{Fig1}, the
metastable fluid-fluid and the stable fluid-solid phase lines taken
from simulations\cite{Dijk98} are compared to lines of constant
coordination number, which are calculated with eqn.~(\ref{eq1}) and
$g_{ll}(r) = \exp \left( - \beta V_{dep}(r) \right)$, together with the
depletion potential used for the simulations as well the simpler AO
potential.  The difference between the two is very small, implying
that the coordination number is not very sensitive to the exact form
of the depletion potential.  Also included is the proposed
bond-percolation induced fluid-solid line at $n_b=2.4$ derived from
the approximation to $g_{ll}(r)$ used by Buhot.  
 \vglue - 0.3cm
\vglue - 0.5cm

   Clearly: {\bf (a)} As expected, the approximation used by Buhot for
$g_{ll}(r)$ breaks down as $\eta_l$ increases.
 {\bf (b)} The lines of constant coordination number are not related
to either the fluid-fluid or the fluid-solid phase lines, implying
that there is no direct relation between bond-percolation and
phase-separation.  The same results were found for other size ratios,
and
it is hard to see how more accurate approximations for $g_{ll}(r)$
could  change this picture.

The breakdown of the bond-percolation picture for this archetypical
hard-core mixture model implies a similar breakdown for other, more
complex, mixtures.  The good agreement found\cite{Buho99} at a few
state points for hard-square systems probably results from either
their 2-d nature, the imposed parallel symmetry, or the rather
unusual purported 2nd order fluid-solid transition. It does not imply
that bond-percolation is generally relevant for fluid-solid
phase-separation in binary hard-core mixtures.

\small 

\noindent A.A. Louis

	Department of Chemistry, Lensfield Rd,

        Cambridge CB2 1EW, UK \vspace{5pt}


\end{multicols}

\end{document}